\documentclass[12pt]{article}
\usepackage{authblk}
\usepackage{amsmath}
\usepackage{float}

\usepackage{graphicx}
\usepackage{dcolumn}
\usepackage{bm}

\usepackage[natbib=true,style=numeric,sorting=none]{biblatex}
\usepackage[right=2.5cm, bottom=2.5cm, left=2.5cm, top=2cm]{geometry}

\usepackage{xcolor}

\usepackage{lineno}

\begin{document}

\title{Design and characterization of a fast output amplifier for Silicon Photomultipliers}

\author[1]{\small Felipe Soriano\footnote{Lead author: fsoriano@estudiantes.unsam.edu.ar}}
\author[2]{Lucas Finazzi\footnote{Lead author/corresponding author: lfinazzi@unsam.edu.ar}}
\author[2]{Gabriel Sanca}
\author[2]{Federico Golmar}

\affil[1]{Escuela de Ciencia y Tecnología, Universidad de San Martin, Buenos Aires, Argentina}
\affil[2]{Instituto de Ciencias Físicas, Universidad de San Martin, CONICET, Buenos Aires, Argentina}

\date{\today}

\maketitle

\begin{abstract}
    A voltage amplifier, based on the BFU500XRR NPN transistor in a common-emitter configuration, was developed for the readout of the fast output of an ONSEMI MicroFC-10035 Silicon Photomultiplier (SiPM). This amplifier was tested and characterized under dark and illuminated SiPM conditions. For the design presented in this work, a Gain of $(20.0 \pm 0.7)$~dB and a rise time of $(768 \pm 10)$~ps were achieved for one amplifier stage and a Gain of $(38.3 \pm 0.7)$~dB and rise time of $(1155 \pm 16)$~ps was achieved for two cascaded stages. In addition, the linearity of both topologies was verified and the measured Signal-to-Noise Ratio (SNR) was ($21.9 \pm 0.4$) dB and ($18.8 \pm 0.5$) dB for the one-stage and two-stage amplifiers, respectively. This design provides a cheap and scalable (weight and size) alternative to other fast output amplifiers that are used when operating SiPMs fast outputs. These characteristics make this design ideal for space applications.
\end{abstract}

\vspace{5pt}

\section{Introduction}
\label{sec:intro}

Silicon photomultipliers (SiPMs) are advanced solid-state optoelectronic devices that are compact, durable, immune to magnetic fields, and require relatively low bias voltage to operate ($\sim$30~V). They are also very sensitive, and have the ability to detect single-photons at room temperature~\cite{sipm_review}. They consist of a dense array of Single Photon Avalanche Diodes (SPADs) operated in Geiger mode, which can be triggered by incoming photons individually. When an incoming photon is detected, an avalanche is triggered in the device and a macroscopic signal can be read out from one of its terminals. The output generated depends on the amount of SPADs triggered by incoming photon events, which means that SiPMs have the capability to measure the number of incoming photons at any given time~\cite{sipm_review2} (one triggered SPAD is often referred as a 1 p.e. event, two SPADS as a 2 p.e. even, etc). This makes them more versatile than other single-photon detection devices, like APDs, for certain applications where a photon number measurement is needed. SiPMs have been used in a myriad of different applications, including quantum optics~\cite{lf_bunching}, medical imaging~\cite{pet1, pet2}, high energy physics~\cite{hep1, hep2}, space applications~\cite{sipm_space1, sipm_space2}, optical communications~\cite{comm1, comm2}, LIDAR~\cite{lidar1}, among others.

Various commercial SiPMs have two output terminals: A standard output and a fast output. The fast output terminal delivers an ultra-fast voltage pulse output signal with a sub nanosecond rise time and a Full Width at Half Maximum (FWHM) approximately between 0.3~ns and 0.6~ns. The availability of two separate outputs is particularly significant in applications requiring simultaneous measurement of both the timing and energy of incoming photon events. The fast-output signal is crucial in scenarios where precise timing is paramount.

In this work, a voltage amplifier was developed for the fast output of an ONSEMI MicroFC-10035 SiPM using a single BFU500XRR transistor per amplifier stage. The design was tested in a single-stage and dual-stage configuration and the results are presented here. This design is planned for integration with a previously designed amplifier~\cite{afe_standard} for the SiPM standard output. Both amplifiers are planned to be used in the context of visual light communication between Earth and a satellite in Low Earth Orbit. The group has extensive experience in operating LabOSat platforms in such space settings~\cite{lf_begonia, lf_cae2023, 9_missions_decade, lf_cae2024, lf_dosims}. 

In Section~\ref{sec:design}, the schematic for each stage is detailed and design consideration are discussed. In Section~\ref{sec:setup}, the experimental setup to test these amplifiers is described. In Section~\ref{sec:results}, the measurement results are presented and discussed. In Section~\ref{sec:conclusions}, the conclusions are presented and future work is outlined.

\section{Electronics Design}

\label{sec:design}

The preliminary requirements for the amplifier are presented in Table~\ref{tab:reqs}.

\begin{table}[H]
    \centering
    
    \begin{tabular}{|c|c|}
        \hline
         \textbf{Parameter} & \textbf{Design requirement} \\ 
        \hline
        Gain & 20~dB \\ 
        \hline
        Rise Time & 750~ps \\ 
        \hline
        Max. power & 200~mW \\
        \hline
    \end{tabular}
    \caption{Preliminary requirements for every stage of the designed SiPM fast output amplifier.}
    \label{tab:reqs}
\end{table}

The Gain requirement is due to the fact that SiPM fast output pulses are expected to be as low as $\sim$1~mV, so the two-stage amplifier output was desired to be 100~mV (40~dB Gain). This output requirement value was set so the amplifier output could be digitized properly with commercial on-chip ADCs with 12-bit resolution. The bandwidth was chosen based on the SiPM manufacturer's estimation of the fast output rise time. The power requirement comes from the fact that this amplifier is planned to be used in space, where power is often limited. Also, the input and output impedance was chosen as 50~$\Omega$ for compatibility with traditional measurement instruments. With matching input and output impedances of 50~$\Omega$, this amplifier can theoretically be cascaded multiple times with minimal signal reflection. 

\begin{figure}[H]
\centering
    \includegraphics[width=0.7\textwidth]{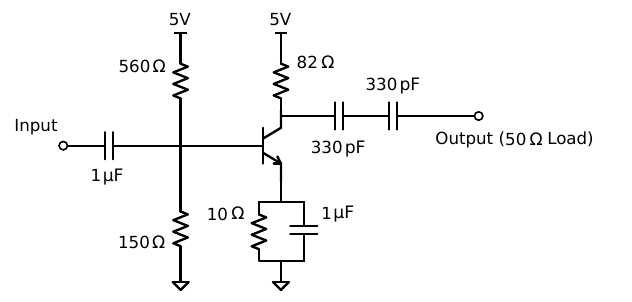}
    \caption{Schematic of the one-stage amplifier design based on the BFU500XRR NPN transistor. This design can be cascaded multiple times to increase Gain with minimal input/output signal reflections.}
    \label{fig:circuit}
\end{figure}

The one-stage fast output amplifier designed based on the requirements presented in Table~\ref{tab:reqs} is shown in Figure~\ref{fig:circuit}.

As mentioned, when working with the fast-output, achieving high timing resolution is critical. For this reason, minimizing both rise time and recovery time was a priority. The target Gain was set at 20~dB per stage, while maintaining the pulse shape as close to the original as possible. To meet these requirements, a transistor with a flat Gain up to 1~GHz was selected. The BFU550XRR NPN transistor was chosen because it provides the necessary Gain along with a low noise figure. In addition, its widespread availability and cheap cost were a big upside in this selection.


It is worth highlighting that the fast-output of the SiPM is a voltage pulse, which does not contain a specific frequency component but rather a continuous spectrum of frequencies. As a result, designing a matching circuit for the 50~$\Omega$ output of the SiPM evaluation board using traditional reactive components or common matching configurations presents significant challenges, particularly when balancing the trade-offs between impedance matching, Gain, and, above all, stability. Nevertheless, impedance matching remains necessary. A suitable approach was found by biasing the transistor's with a voltage divider. From a small-signal perspective, the circuit's input impedance can be modeled as a parallel network consisting of three components: the 150~$\Omega$ and 560~$\Omega$ resistors that form the voltage divider, and the impedance presented by the transistor at the base. The latter depends, among other factors, on the transistor current gain $h_{\mathrm{FE}}$, with lower values of $h_{\mathrm{FE}}$ leading to a reduced input resistance. The transistor in the circuit operates with a $h_{\mathrm{FE}}$ of 83, and careful selection of the values of the voltage divider resistors ensures that the reflections at the input remain below -10~dB across the operating frequency spectrum.

The output was matched using the same approach. The collector resistance of 82~$\Omega$ was chosen so that the parallel network, formed with the impedance when looking into the transistor’s collector, closely aligned with the load impedance requirement. Since the circuit’s Gain depends on the value of this resistor, a trade-off must be made between achieving optimal Gain and minimizing signal reflections to ensure proper impedance matching. Finally, a 10~$\Omega$ resistor was added to the emitter to enhance the circuit's stability, with a parallel capacitor included to prevent further reduction in Gain. One viable alternative for impedance matching could have been balun transformers, which was already used in the bibliography~\cite{Cates_2022} with successful results.

The circuit was manufactured in-house on a two-layer PCB with an FR4 core of 1.6~mm thickness. The top layer contained all traces, with the remaining copper used to establish a ground plane. The bottom layer served as an additional ground plane, allowing the signal-carrying traces to be designed as coplanar waveguides. This ensured the impedance stayed within a 5~\% of the 50~$\Omega$ design target. Two 50~$\Omega$ Amphenol SMA connectors were used for both input and output signals.

Component placement and routing were carefully planned to preserve signal integrity. Traces were kept as short and direct as possible, with a surrounding ground plane to provide a proper return path for the signal. To minimize crosstalk, traces not carrying the signal were positioned perpendicularly to those that did. No special thermal considerations were necessary due to the circuit's low power consumption. Although parasitic effects were considered during design, no additional simulations were performed to model these parasitics. With the exception of the SMA connectors, all components were soldered using a reflow oven.

\section{Measurement Setup}

\label{sec:setup}

    The tests were performed using an ONSEMI MicroFC-10035 SiPM, which has both a standard and a fast output. The SiPM itself was biased with 30~V using a Keithley 2612B SourceMeter Unit. The fast output of this SiPM was connected to the one-stage and two-stage amplifiers to be tested. A 5 Series MSO54 Tektronix Oscilloscope (2~GHz bandwidth) was used to acquire the direct SiPM's fast output waveform and the amplifier output waveforms simultaneously. The raw fast output pulse was used as the input of a SPICE simulation and both the simulated and real amplifier outputs were compared for the same input pulse. The power consumption of one-stage was 110 mW, which is below the 200~mW target. This means that the amplifier can be integrated with other low power designs for space applications. All tests were performed in a dark room to avoid external light stimuli that could have compromised the measurement. 
    
    For each configuration, several parameters were studied:
    
    \begin{itemize}
        \item \textbf{Gain:} Calculated from the slope of a linear fit on the plot of SiPM pulse maximum voltage vs. amplifier output pulse maximum voltage.
        \item \textbf{Rise Time:} Calculated from the acquired waveforms as the time it takes the signal to go from 10~\% to 90~\% of its maximum amplitude.
        \item \textbf{Pulse FWHM:} Calculated as the width of the pulse at half its maximum pulse amplitude.
        \item \textbf{Signal-to-Noise Ratio (SNR):} Calculated as the ratio between a 1 p.e. pulse maximum amplitude vs. the corresponding baseline standard deviation. 
    \end{itemize}
    
\section{Results and Discussion}

\label{sec:results}

\begin{figure}[H]
    \centering
    \includegraphics[width=0.7\linewidth]{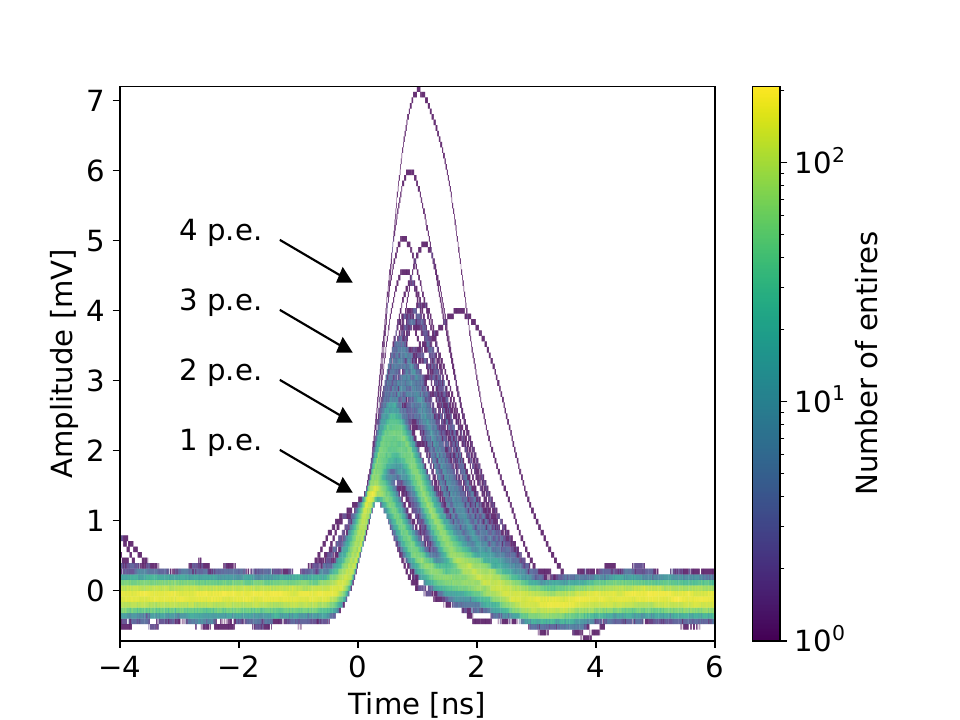}
    \caption{Persistence histogram including all acquired SiPM fast output events. It can be seen that the voltage output of the SiPM varies depending on the amount of SPADs triggered (i.e. 1 p.e. means one triggered SPAD, 2 p.e. means two triggered SPADs, etc). This gives the SiPM the ability to detect many incoming photons at a given time. The smallest pulse is approximately $(0.8 \pm 0.1)$~mV. The sub-nanosecond rise time can also be observed for all events.}
    \label{fig:persistence_hist}
\end{figure}

In this section, the results obtained are presented. For reference, a persistence histogram of the SiPM's fast output can be seen in Figure~\ref{fig:persistence_hist}.

The output voltage of the SiPM varies depending on the amount of SPADs triggered (i.e. 1 p.e. means one triggered SPAD, 2 p.e. means two triggered SPADs, etc). This gives the SiPM the ability to detect many incoming photons at a given time. The smallest pulse has a voltage of $(0.8 \pm 0.1)$~mV. For the raw SiPM output pulse, the measured 1 p.e. rise time was $(750 \pm 8)$~ps, the 1 p.e. pulse FWHM was $(1.18 \pm 0.01)$~ns and the SNR for the 1 p.e. pulse was $(15.9 \pm 1.4)$~dB.

\begin{figure}[H]
\centering
    \includegraphics[width=0.7\textwidth]{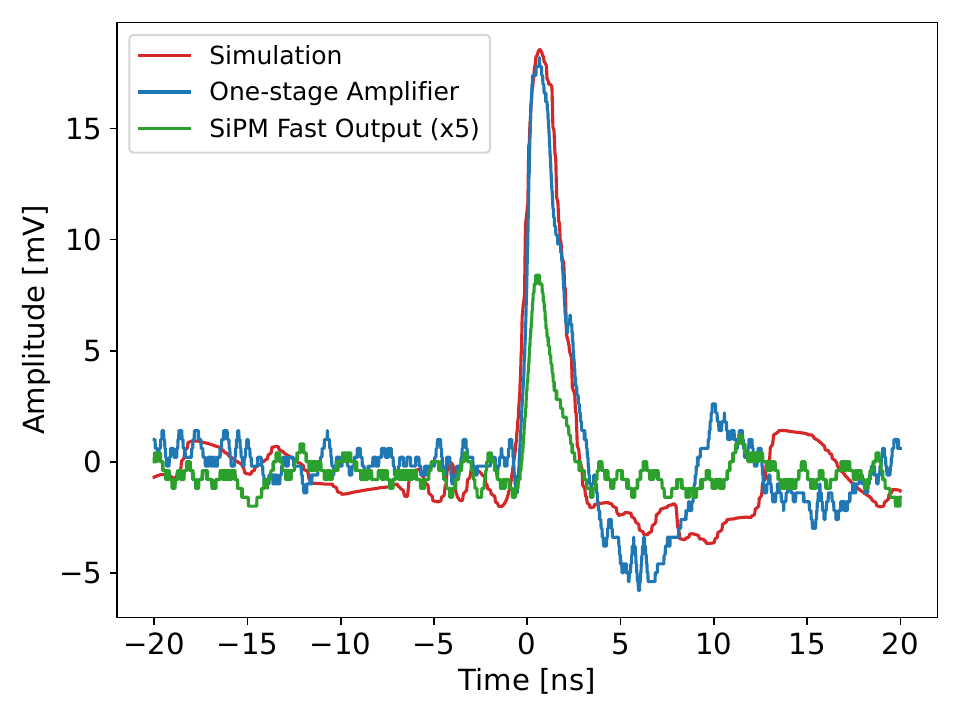}
    \caption{Plot of measured signal as a function of time (uncertainties are omitted for plot clarity). The original pulse of the fast output is shown, along with the simulated and measured one-stage design. The SiPM fast output is multiplied by 5 for better curve comparison. The amplifier output has a measured Gain of $(20.0 \pm 0.7)$~dB, which is compatible with the desired target Gain of 20~dB. The pulse undershoot was a design compromise to reduce the amplifier’s settling time. Although the measured undershoot is larger than expected, the recovery time remains within the acceptable margin.}
    \label{fig:sim_comp}
\end{figure}

In addition, a comparison between the SPICE simulation and the one-stage amplifier was performed (for the same SiPM input pulse). This comparison is shown in Figure~\ref{fig:sim_comp}.

\begin{figure}[!ht]
    \centering
    \includegraphics[width=0.7\linewidth]{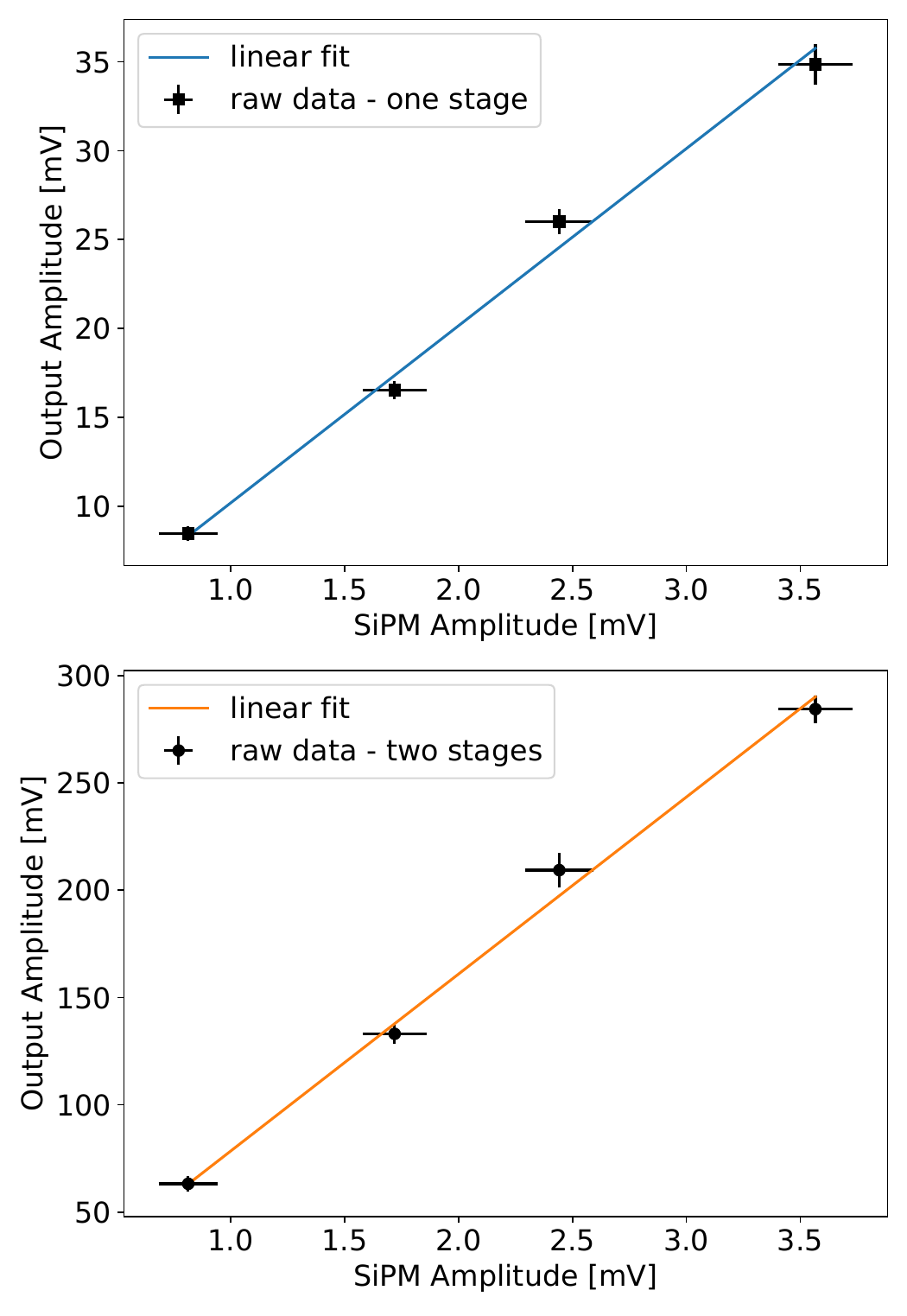}
    \caption{Amplifier linearity for the one and two-stage variants with corresponding linear fits. Both the one and the two-stage versions show linearity for all observed SiPM fast output pulses. The slope for these linear fits give the Gain of the one-stage and two-stage version of the amplifier. P-values for both fits give 0.52 and 0.62, for the one-stage and two-stage variants, respectively. This indicates a satisfactory goodness of fit and a high linearity.}
    \label{fig:linearity}
\end{figure}

The amplifier output has a measured Gain of $(20.0 \pm 0.7)$~dB, which is compatible with the desired target Gain of 20~dB. The Gain was calculated with a linear fit of a plot of output signal vs. input SiPM signal, which is shown in Figure~\ref{fig:linearity}. The observed undershoot is a design trade-off, necessary in order to shorten the amplifier's settling time. The higher undershoot seen in the implemented amplifier could be due to impedance mismatches and other parasitic capacitances and inductances that add non minimum phase zeros. The rise time was $(768 \pm 16)$~ps, which is compatible with the original SiPM rise time, showing no rise time degradation. In addition, the FWHM of the pulse was $(1.88 \pm 0.01)$~ns and the Signal-to-Noise Ratio was $(21.9 \pm 0.4)$~dB. This SNR is larger than the SiPM output, which means that the signal amplitude is amplified more than the baseline noise, which is desired in any voltage amplifier.

\begin{figure}[H]
\centering
    \includegraphics[width=0.7\textwidth]{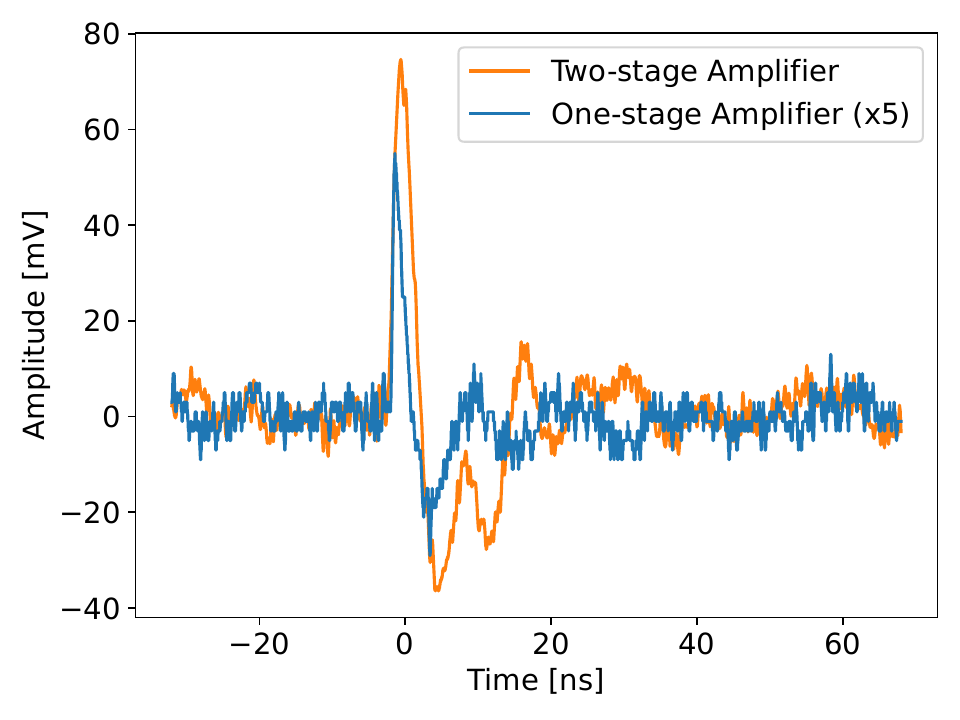}
    \caption{Plot of amplifier outputs as a function of time. The amplitude of the one-stage amplifier is multiplied by 5 for better curve comparison. A total Gain of $(38.3 \pm 0.7)$~dB was measured for the two-stage version. Some oscillations can be seen with the two stages in the pulse's undershoot. These oscillations make it so that the measured Gain is slightly smaller than the target 40~dB. Nevertheless, the measured Gain value and the target value are compatible with one another.}
    \label{fig:stage_comp}
\end{figure}

In addition, two one-stage amplifiers were cascaded to make a two-stage amplifier. The performance against the one-stage variant is shown in Figure~\ref{fig:stage_comp}.  

The amplitude of the one-stage amplifier is multiplied by 5 for better curve comparison. A total Gain of $(38.3 \pm 0.7)$~dB was measured for the two-stage version. Some oscillations can be seen with the two stages in the pulse's undershoot. These oscillations can be due to imperfections in the impedance matching between the first amplifier output and the second amplifier input. These oscillations make it so that the measured Gain is slightly smaller than the target 40~dB. To reduce these oscillations, the two stages could be added to the same PCB board to have a better impedance control over the required trace. In contrast with the one-stage variant, the rise time for the two stage version is $(1155 \pm 16)$~ps. The rise time is degraded by a factor of 1.5. Again, this could be due to reflections between the two amplifiers. In addition, the FWHM of the pulse was $(2.63 \pm 0.02)$~ns and the Signal-to-Noise Ratio was $(18.8 \pm 0.5)$~dB. This value is slightly smaller than the one-stage value and is caused by the amplification of the noise from the first stage. 

The amplifier linearity was also studied for the one and two-stage variants. The results are shown in Figure~\ref{fig:linearity} and the slope of these plots gives the Gain reported previously for the one-stage and two-stage amplifiers, respectively.

Both the one and the two-stage versions show linearity for all observed SiPM fast output pulses. To observe 4 p.e. events, the SiPMs had to be illuminated during the measurements. 

A brief summary of the measured parameters of interest are reported in Table~\ref{tab:meas_params}.

\begin{table*}[!t]
    \renewcommand{\arraystretch}{1.3}
    \label{tab:meas_params}
    \centering
    \resizebox{\columnwidth}{!}{%
    \begin{tabular}{|c|c|c|c|}
        \hline
         \textbf{Parameter} & \textbf{Measurement (raw SiPM output)} & \textbf{Measurement (one-stage)} & \textbf{Measurement (two-stage)} \\ 
        \hline
        Gain [dB] & - & $20.0 \pm 0.7$ & $38.3 \pm 0.7$ \\ 
        \hline 
        Rise time [ps] & $750 \pm 8$ & $768 \pm 10$ & $1155 \pm 16$ \\ 
        \hline
        Pulse FWHM [ns] & $1.18 \pm 0.01$ & $1.88 \pm 0.01$ & $2.63 \pm 0.02$ \\
        \hline
        SNR [dB] & $15.9 \pm 1.4$ & $21.9 \pm 0.4$ & $18.8 \pm 0.5$\\ 
        \hline
    \end{tabular}
    }
    \caption{Measured parameters and comparison for the SiPM raw output, one-stage amplifier output and two-stage amplifier output.}
\end{table*}

\begin{figure}[!ht]
\centering
    \includegraphics[width=0.7\textwidth]{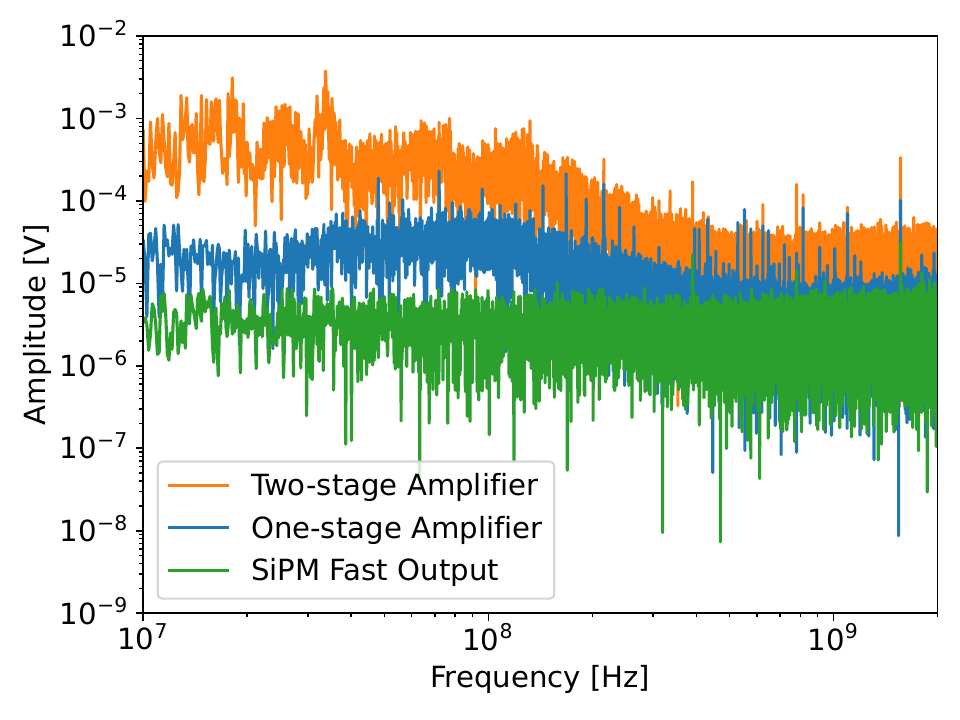}
    \caption{FFT comparison of the raw SiPM output pulse and the outputs after one-stage and two-stage amplification. Both amplifier stages show reduced high-frequency components, with the two-stage amplifier exhibiting a more pronounced roll-off. While the overall frequency content is preserved, the sharper attenuation in the two-stage configuration is consistent with the expected behavior when cascading amplifiers.}
    \label{fig:fft_comparison}
\end{figure}

Additionally, an FFT of the acquired signals was performed to observe what spectral components the amplifiers add to the output signal when compared to the SiPM's raw fast output pulse. A comparison of all the FFTs can be seen in Figure~\ref{fig:fft_comparison}. 

Both amplifier stages show reduced high-frequency components; however, the two-stage amplifier exhibits a more pronounced roll-off. This sharper attenuation can be attributed to the roll-off multiplication effect that occurs when amplifiers are cascaded, where each stage contributes additional poles to the transfer function. Moreover, variations in component tolerances and fabrication factors may introduce parasitic elements that further impact high-frequency performance. While the overall frequency content is preserved, these factors collectively contribute to the greater loss of high-frequency detail in the two-stage configuration, consistent with the expected behavior when cascading amplifiers.

\section{Conclusions and outlook}

\label{sec:conclusions}

In this work, the design and characterization of a voltage amplifier, utilizing the BFU500XRR NPN transistor in a common emitter configuration, were presented. This amplifier was developed for the readout of the fast output of an ONSEMI MicroFC-10035 SiPM and its use in space applications. For the one-stage amplifier, the measured Gain was $(20.0 \pm 0.7)$~dB and the rise time was $(768 \pm 10)$~ps. Furthermore, the two-stage amplifier had a measured Gain of $(38.3 \pm 0.7)$~dB and the rise time of $(1155 \pm 16)$~ps. One stage of the amplifier doesn't degrade the rise time of raw SiPM fast output pulses. Nevertheless, the two-stage cascaded version degrades the rise time by a factor of 1.5. This could be due to the slight impedance mismatch between the first amplifier's output and the second amplifier's input. The linearity of one and two stages was also verified and the fit slopes (or the Gain) for both stages and the SNR for was ($21.9 \pm 0.4$) dB and ($18.8 \pm 0.5$) dB for the one-stage and two-stage amplifiers, respectively.

Further tests will be performed in the future by placing both amplifier stages in the same PCB. This will allow for better impedance control between the first and second stage amplifiers. This is expected to improve the observed oscillations and decrease the measured rise time degradation. 

Furthermore, a new board design will also be developed. This design will have two stages based on the BFU500XRR. The topology of the first stage will be retained as presented in this work; however, a self-biasing configuration will be employed for the second stage. By maintaining the same impedance matching strategy, it is sufficient to modify the values of the passive components to potentially enhance the performance of the cascaded two-stage amplifier. This new design will be combined with an SiPM standard output amplifier in a single board and will be used for visual light communication in Low Earth Orbit using LabOSat platforms.

\section*{Acknowledgements}

The authors acknowledge financial support from \mbox{ANPCyT PICT 2017-0984} ``Componentes Electrónicos para Aplicaciones Satelitales (CEpAS)'', \mbox{PICT-2019-2019-02993} \mbox{``LabOSat:} desarrollo de un Instrumento detector de fotones individuales para aplicaciones espaciales'' and \mbox{UNSAM-ECyT} FP-001.

\printbibliography

\end{document}